\documentclass[11pt,conference]{IEEEtran}
 \IEEEoverridecommandlockouts

\pdfoutput=1
\usepackage{graphicx}

\newcommand{\bgref}{\mbox{$^\star$}}

\title{A Preliminary Review of Influential Works in Data-Driven Discovery}

\author{
\IEEEauthorblockN{Mark Stalzer and Chris Mentzel}
\IEEEauthorblockA{Science Program\\
Gordon and Betty Moore Foundation\\
Palo Alto, California 94304\\
stalzer at caltech.edu and chris.mentzel at moore.org}
}

\begin{document}

\maketitle

\begin{abstract}
The Gordon and Betty Moore Foundation ran an Investigator Competition as part of its Data-Driven Discovery Initiative in 2014. We received about 1,100 applications and each applicant had the opportunity to list up to five {\em influential works} in the general field of ``Big Data'' for scientific discovery. We collected nearly 5,000 references and 53 works were cited at least six times. This paper contains our preliminary findings.
\end{abstract}

\section{Introduction}
The long-term goal of the Gordon and Betty Moore Foundation's Data-Driven Discovery Initiative (DDD) is to foster and advance the people and practices of data-intensive science to take advantage of the increasing volume, velocity, and variety of scientific data to make new discoveries. Data-intensive science is inherently multidisciplinary, combining natural sciences with methods from statistics and computer science.

In January 2014 the DDD launched an Investigator Competition (IC) to identify some of the leading innovators in data-driven discovery. These scientists are striking out in new directions and are willing to take risks with the potential of huge payoffs in some aspect of data-intensive science. As part of the competition we collected several thousand references, which we call {\it influential works}, to the literature, software, and data sets that the applicants listed as one of the top five most important works in data-intensive science or data science.

This paper is a preliminary review of what we found. The next section presents the methodology and some statistics from the references. Section~\ref{sec:clusters} contains several natural clusters of the works, some are obvious like genomics and machine learning. Others like the impact of Google's work, and questions about the scientific method are perhaps of more general interest. This paper ends with some limitations and next steps.

\section{Influential Works at a Top Level}

In the competition pre-application stage we asked for up to five {\em influential works} in data-driven discovery. Specifically, as stated in the competition FAQ:

\begin{quotation}
The (up to) five Influential Works on the pre-application web form are for you to reference 
work that you think has helped define the field of data science. This may or may not be your 
own work. Taken collectively, across all the DDD IC pre-applications, these works will give the 
foundation a snapshot of data intensive science.
\end{quotation}

A total of 1,095 applications were received in late February 2014, containing 4,790 references.

The raw data is not available for public release since it was collected with the Foundation's promise of anonymity to get a better sampling. Specifically, from the competition FAQ:
\begin{quotation}
  Members of the DDD staff intend to write a review paper that summarizes these findings, and information will only be used in an aggregate form.
\end{quotation}
Presented in this paper is an aggregate form, via an automated sorting process that is described in the Appendix, for works cited at least six times. There are 53 of these works; and the ones cited at least ten times are in Table~\ref{tab:topworks}. This automatic approach works very well for papers and books, which have a well established citation form, but not so well for resources and tools and this will be discussed further in the limitations part of the concluding remarks.

\begin{table}[t]
\centering
\begin{tabular}{l | l | l}
Count & Year & Citation \\
\hline
63 & 2008 & MapReduce\cite{dean2008mapreduce} \\
51 & 2009 & {\it Fourth Paradigm}\cite{hey2009fourth} \\
43 & 2009 & {\it Elements of Statistical Learning}\cite{hastie2009elements} \\
30 & 2001 & Initial sequencing of the human genome\cite{lander2001initial} \\
24 & 1948 & A mathematical theory of communication\cite{shannon2001mathematical} \\
23 & 2000 & Sloan Digital Sky Survey\cite{york2000sloan} \\
20 & 1990 & BLAST\cite{altschul1990basic} \\
19 & 1996 & Lasso\cite{tibshirani1996regression} \\
19 & 2003 & Latent Dirichlet allocation\cite{blei2003latent} \\
17 & 1977 & EM algorithm\cite{dempster1977maximum} \\
17 & 1995 & Support vector networks\cite{cortes1995support} \\
15 & 2001 & Random forests\cite{breiman2001random} \\
14 & 2006 & {\it Pattern Recognition}\cite{bishop2006pattern} \\
14 & 1998 & Anatomy of web search engine\cite{brin1998anatomy} \\
13 & 2007 & {\it Numerical Recipes}\cite{press2007numerical} \\
11 & 1979 & Bootstrap methods\cite{efron1979bootstrap} \\
11 & 1953 & Equation of state calculations\cite{metropolis1953equation} \\
11 & 1977 & Exploratory data analysis\cite{tukey1977exploratory} \\
11 & 1988 & {\it Probabilistic reasoning}\cite{pearl1988probabilistic} \\
10 & 1999 & PageRank\cite{page1999pagerank} \\
10 & 2013 & {\it Bayesian Data Analysis}\cite{gelman2013bayesian} \\
10 & 2009 & Unreasonable effectiveness of data\cite{halevy2009unreasonable}
\end{tabular}
\smallskip
\caption{Works that were cited at least ten times, with count, year, and citation.}
\label{tab:topworks}
\end{table}

\begin{figure}[t]
\centering
\includegraphics[scale=0.4]{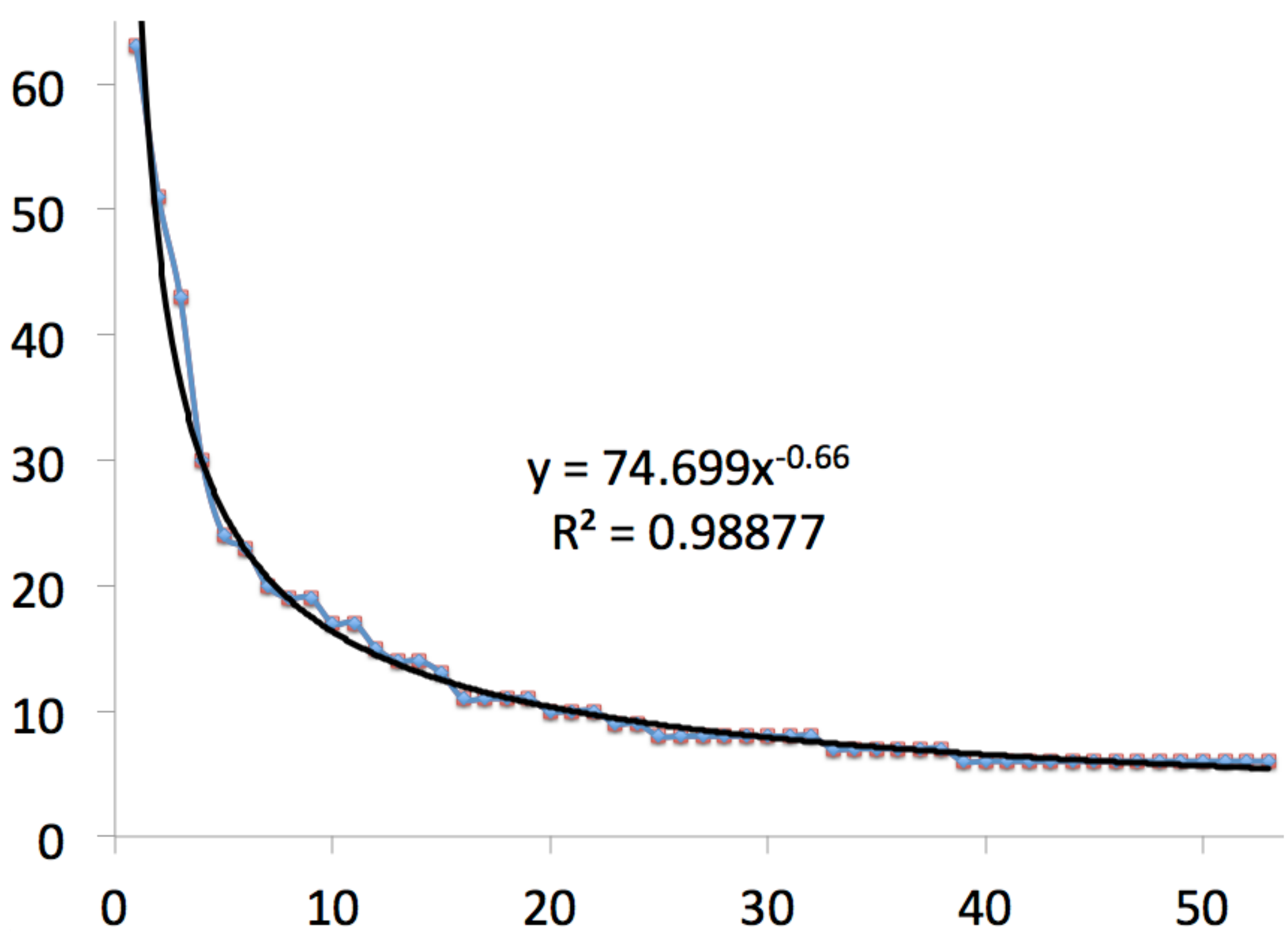}
\caption{Fit of the influential works to a power law (x is index, y is count). The correlation coefficient is $R^2 = 0.989.$}
\label{fig:power}
\end{figure}

A plot of the reference index for all works versus the reference count fitted to a power law is shown in Figure~\ref{fig:power}. The correlation of about 0.99 is very good agreement. The h-index of the works is 14; this is the subset of the works cited as least as often as their rank by the number of times cited\cite{hirsch2005index}\bgref\footnote{References that provide background information, and not in the 53 influential works found as part of the competition, are denoted by a $\star$.}.

The data set is 1.7MB and is difficult to examine directly, but the sorting processing was manually validated on some references that have rare words. For example, MapReduce\cite{dean2008mapreduce} is reported here with 63 citations and a hand count shows 64, Latent Dirichlet allocation\cite{blei2003latent} is a perfect 19 for 19, and The Fourth Paradigm\cite{hey2009fourth} is 51 for 58 and this mostly was due to sloppy citations. The counts reported here can be considered good lower bounds on the real counts.

\section{Clusters of Influential Works}
\label{sec:clusters}

The works were manually organized into clusters by natural science domain, methodologies, tools, and the scientific method as shown in Table~\ref{tab:clusters}. Each cluster has some key topics as described below and all influential works are cited with varying levels of description.

\begin{table}[h]
\centering
\begin{tabular}{l | l | l l}
Count & Cluster & Key Topics \\
\hline
7 & Domain Sciences (\ref{sec:domains}) & Astronomy \\
  & & Genomics \\
29 & Methodologies (\ref{sec:meths}) & Theory \\
  & & Statistical Methods \\
  & & Machine Learning \\
9 & Tools and Apps (\ref{sec:tools}) & Google \\
  & & General Tools \\
8 &  Scientific Method (\ref{sec:scimeth}) \\
\hline
53 & ALL
\end{tabular}
\smallskip
\caption{A clustering of the 53 influential works with associated section numbers.}
\label{tab:clusters}
\end{table}

\subsection{Domain Sciences}
\label{sec:domains}

\subsubsection{Astronomy}

The Sloan Digital Sky Survey (SDSS)\cite{york2000sloan} is a widely cited resource (\verb|www.sdss.org|)\footnote{SDSS was cited as both a resource and an associated technical summary paper. The intent was clear so we grouped all the citations together.}. The current release is SDSS-III DR12 that has observations through 14 July 2014 and contains 469,053,874 unique, primary, sources from several datasets. Generally, online astronomical datasets are being federated via interoperability standards created by organizations such as the International Virtual Observatory Alliance (\verb|ivoa.net|). The result is a virtual telescope, and astronomers have been pioneers in making observations openly available and accessible.

New instruments are also showing that data-driven discovery is not just about the volume of data, but also the ``velocity''. One of the major challenges with the Large Synoptic Survey Telescope (LLST), which should start doing science runs about 2020, is that the number of alerts to interesting objects may overwhelm the available follow up resources. Good object classification (\ref{sec:ml}) and prioritization will be crucial to the science output. 

\subsubsection{Genomics}

It is very clear that genomics and the Human Genome Project (HGP) have been the main driver of data driven discovery in the life sciences. The two primary works are the ``Initial sequencing and analysis of the Human genome''\cite{lander2001initial}  and the related paper by Venter et al.\cite{venter2001sequence}. These papers report the sequencing of the approximately 3 billion nucleotides that make up the human genome. The project was considered essentially complete in April 2003 and according to the NIH's HGP factsheet, it has enabled the discovery of over 1,800 disease related genes and many other applications. An example is the Thousand Genomes Project\cite{10002012integrated} which, as of 2012, had completed a variety of sequences from 1,092 individuals from 14 populations. This allows comparative analysis of the sequences, which is at the core of bioinfomatics-based discovery.

Consider two sequences $a, b$ composed from the alphabet $\{A, C, G, T\}$ -- DNA nucleotides. We want to find the optimal alignments, essentially a string matching problem, of $a, b.$ In general, however, the alignments are not perfect string matches due to missing data and other factors. Instead, a distance metric is defined and the alignments are optimized with respect to that metric. For example, under a certain metric two good alignments of {\tt GACTAC} are {\tt -ACG-C} and {\tt -AC-GC}. This can be done optimally using dynamic programming in time $O(|a| |b|).$ However, if $a$ must be aligned with many $b$ taken from a database search, the computational expense is prohibitive. A key bioinformatics tool is the ``Basic Local Alignment Search Tool''\cite{altschul1990basic} (BLAST). BLAST uses heuristics to reduce the time complexity and make large-scale searches practical.

There are many other applications besides human health. For example, population groupings can be inferred using Bayesian clustering methods from {\it multiloci} genotype information\cite{pritchard2000inference}. This is an early form of Latent Dirichlet allocation (LDA) which is described more fully in the next section. It can be thought of as running LDA on genetic data, rather than on text: it clusters {\em individuals} into {\em population} rather than documents into topics\footnote{The method can be used back in time since DNA can be preserved; population studies have been done on Darwin's finches from the Gal\'apagos in 1835 using specimens from British museums\cite{petren2010multilocus}\bgref.}. Another emerging example is the use of bioinformatics methods in ecology\cite{jones2006new}. A major challenge here is the heterogeneous natures of the data, from individuals to the biosphere, and their interactions.

The Protein Data Bank\cite{berman2000protein} was established at Brookhaven National Lab in 1971 as an archive for structural genomics data: essentially the shapes of biologically active molecules. These shapes and other information is determined experimentally by X-ray diffraction, NMR, and sometimes theoretical modeling. These experiments require special facilities and can be costly, so there was clearly a motivation in the community to build an archive to minimize duplication of effort. In 2000 there were 10,714 structures and this has grown to 106,710 by early 2015. The data bank supports sophisticated query mechanisms to assist researchers in finding structures with certain properties, such as atomic locations.

It is interesting that the two most referenced natural science domains are astronomy and genomics, and they can differ in length scales of phenomena by up to 33 orders of magnitude. The fact that humanity can probe over such a large range, and even further with high-energy physics experiments, is simply amazing.

\subsection{Methodologies}
\label{sec:meths}

\subsubsection{Foundational Theory}

Reverend Bayes' essay on the Doctrine of Chances in 1763\cite{mr1763essay} is the earliest commonly cited paper and it is truly foundational for data science (a popular modern text is Gelman et al.\cite{gelman2013bayesian}). The work introduces ``Bayes Law'' which gives the likelihood of a condition $A$ being present given that condition $B$ is present, denoted as the conditional probably $P(A | B)$, as
\begin{equation}
P(A | B) = {P(B |A) P(A) \over P(B)}
\label{eqn:bayes}
\end{equation}
where $P(A)$ and $P(B)$ are the so called prior probabilities, or the frequencies of occurrence of the conditions. Please note that the wording is careful to not confuse coincidence with causality: the ``law'' is just a statement of an existing closed population. This equation is optimal under a crucial assumption and this can be seen since it is the unique generalization (up to an integration constant) of {\it modus ponens} for probabilistic inference\cite{jaynes2003probability}. The crucial assumption is that the priors are known very well. There are extensions to Equation~\ref{eqn:bayes} known as maximal-entropy methods that are based on ideas from statistical mechanics, i.e.~how much information can be contained in all the possible ensembles of states in a closed system; again Jaynes is a good reference\cite{jaynes2003probability}\footnote{It should be noted this may be a data anomaly as one of the authors cited this work on his homepage. He also cited {\it Sports Illustrated} which may explain random references to sports statistics.}.

Shannon's seminal work on how much information can be transmitted over a communications channel is also based on entropic ideas\cite{shannon2001mathematical}. Recently (2006), Donoho wrote on ``Compressed sensing''\cite{donoho2006compressed}, with an application to image analysis, but the development is a more general result in information theory. Let $x$ be an unknown vector of size $|m|$ and that we plan to make $n$ measurements of $x$ in a variety of ways. It is shown that only $n = O(m^{1/4} \log^{5/2}(m))$ measurements are needed for a bounded error. This is a very interesting result because it shows that with clever measurement, we do not need to collect nearly as much data {\it if} there is an underlying sparse representation of what is being measured (another way to look at this is that there can be a lot of redundancy in representations). As will be seen in Sec.~\ref{sec:ml}, some forms of compression can be automatically learned.

The Metropolis Algorithm\cite{metropolis1953equation} gives a way for sampling large spaces for computing high-dimensional integrals with a bounded convergence rate. {\it Probabilistic reasoning in intelligent systems: Networks of plausible inference}\cite{pearl1988probabilistic} by J.~Pearl also covers Bayesian inference, Bayesian and Markov networks, and more advanced topics of interest to the artificial intelligence community. We suspect that the use of automated reasoning techniques will grow in data science, although there are issues of scalability. Pearl has also written extensively on coincidence and causality.

\subsubsection{Classical Statistical Methods}

Any section on classical statistical methods must begin with linear models of data, such as fitting a line to a set of points using an ordinary least squares (OLS) estimate. The lasso\cite{tibshirani1996regression}, for ``least absolute shrinkage and selection operator,'' can improve on the prediction accuracy of OLS {\it and} also helps with interpretation since it identifies key coefficients in the estimate.

Consider a sample of size $N$, we can certainly compute basic statistics such as the average. With {\it bootstrap} methods\cite{efron1979bootstrap}, the sample is re-sampled multiple times with replacement to generate better statistics, and this is useful with complicated distributions. Extensions to the original 1979 approach use Bayesian methods\cite{rubin1981bayesian}\bgref. This is further developed in ``A decision-theoretic generalization of on-line learning and an application to boosting''\cite{freund1995desicion}. It is an example of combining multiple {\it strategies}, even if they are individually weak, to build robust models. The authors use many example, including betting on horses.

Incomplete data is a very common problem and it can be formalized as follows. Let ${\bf x} \in {\cal X}$ be the {\it complete data} and ${\bf y} \in {\cal Y}$ be the (possibly incomplete) {\it observed data}, and assume there is a mapping such that ${\cal X}({\bf y})$ gives all possible ${\bf x}$ for an observation ${\bf y}$. Given a set of parameters ${\bf \Phi},$ the family of complete sampling distributions $f({\bf x | \Phi})$ is related to the incomplete family $g({\bf y | \Phi})$ by
\begin{equation}
g({\bf y | \Phi}) = \int_{{\cal X}({\bf y})} f({\bf x | \Phi})
\end{equation}
Dempster, Laird, and Rubin present a method for computing maximum likelihood estimates from incomplete data called the {\it EM} Algorithms (for Expectation-Maximization)\cite{dempster1977maximum}; it does this by adjusting the parameters to  maximize $g$ given the observations. The paper has many examples including missing value situations, truncated data, etc. It was read before the Royal Statistical Society and there is extensive commentary in its published form. One comment in particular, by R.~J.~A. Little, is a fine summary: ``Other advantages of the EM approach are (a) because it is stupid, it is safe, (b) it is easy to program, and often allows simple adaptation of complete data methods, and (c) it provides fitted values for missing data.'' An application of the EM algorithm and Bayesian statistics is ``Latent Dirichlet allocation''\cite{blei2003latent} that build a multi-level model for ``collections of discrete data such as text corpora.''\footnote{It would be interesting to apply LDA to the 53 influential works.}

Consider a set of features used to classify objects. A ``random forest''\cite{breiman2001random} is a collection of decision trees where each tree uses some subset ot the features to do a classification; the trees then vote to determine the final class. It is shown that forests are not subject to overtraining, which can be a problem with machine learning methods (see next section).

{\it The Elements of Statistical Learning}\cite[Chapters 3, 10, 8, 15]{hastie2009elements} by Hastie, Tibshirani, and Friedman covers lasso, bootstrap methods, the EM algorithm, and random forests. It is also has chapters on machine learning which is covered in the next section; it is a popular text. An earlier text\cite{breiman1984classification}, also covers regression and tree methods.

When there are multiple hypotheses a standard approach is to control the familywise error rate (FWER) -- closely related to Type I errors. This is a common problem in determining the efficacy of medical procedures. Benjamini and Hochberg suggest\cite{benjamini1995controlling}, instead, to control the number of falsely rejected hypotheses -- the false discovery rate (FDR). FDR can be more powerful when some (null) hypotheses are non-true. 

Isomap\cite{tenenbaum2000global} is an algorithm for reducing the dimensionality of input spaces, e.g.~face recognition. It is broadly applicable whenever non-linear geometry complicates the use of techniques such as Principal Component Analysis (PCA). Another paper on non-linear reduction\cite{roweis2000nonlinear} presents a local, piecewise, linear method for modeling non-linear data. An interesting example is that using PCA on a logarithmic spiral, to first order, just yields a linear fit; yet the curve can be parameterized by its length and maintain its structure.

\subsubsection{Machine Learning}
\label{sec:ml}

Methods for machine learning are crucial for data-driven discovery and are used for both classification and regression analysis. There are several standards texts\cite{mitchell1997machine,duda1999pattern,bishop2006pattern,murphy2012machine}. Here we will focus on two common methods and some recent advances.

Consider the classification problem $f : X \mapsto \{-1, 1\}$ where $X$ is an observation space and $f$ decides if a member of $X$ belongs to one of two categories. For example, in an astronomical image, find all of the quasars with a redshift greater than some value. Machine learning methods take a set of example observations from $X$ and use some {\it generalization} process to build an $f.$

One of the most rigorously-founded ways is to form a ``Support Vector Machine''\cite{vapnik1998statistical,cortes1995support} (SVM). The construction of an SVM attempts to build a hyperplane that divides the examples into the -1 or +1 spaces. In general, the examples are not completely separable and so a kernel $K(x,x')$ is used to project an element $x$ of $X$ into a higher dimensional space where the separation is more complete. It is useful to look at this in more detail, since it clearly shows data, mathematical formulations, and clever algorithms coming together to form an $f.$

A common kernel is $K(x,x') = e^{-(x-x')^2/{2\sigma^2}}$ where $\sigma$ is determined from the data. The selection of a kernel generally requires some insight, particularly when the data is heterogeneous. Consider $l$ training examples ${(x_1, y_1) \ldots (x_l, y_l)}$ where the $y_i \in \{-1, +1\}.$ To construct an SVM, solve the following optimization problem for $\alpha:$
\begin{equation}
  \sum_{i=1}^l \alpha_i - {1 \over 2} \sum_{i,j=1}^l y_i y_j \alpha_i \alpha_j K(x_i, x_j)
\end{equation}
subject to $\sum_{i=1}^l y_i \alpha_i = 0$ and all $\alpha_i \geq 0$. This can be done via quadratic programming which is generally ${\cal NP}$-hard, but due to some constraints in the formulation the optimization can be done quickly using Sequential Minimal Optimization (SMO)\cite{platt1999fast}\bgref. The decision function is then $f(x) = \hbox{sgn} (b + \sum_{i=1}^l y_i \alpha_i K(x, x_i))$ where $b$ is the scalar category separator and can be computed directly given the $\alpha_i$.

Recently SVMs, called SVM+, have been extended to work with an auxiliary ``privileged information'' set $X^\star$ that is {\it available only during classifier construction}\cite{vapnik2009new}\bgref. An example is to use a protein structure prediction code, during training, to help train a classifier. An SVM+ classifier typically performs better than regular SVM. Constructing an SVM+ can also be done fairly quickly using SMO\cite{pechyony2011fast}\bgref. There is an interesting analogy to Shannon's work that is based on the information available in a closed system. With SVM+, the classifier gets trained with access to another system, Vapnik calls it a teacher ($X + X^\star$), and then works independently ($X$) in operation.

Another common classification method are artificial Neural Networks (NN), and the basic ideas go back to 1943\cite{mcculloch1943logical}\bgref. Here the input vector is fed into sigmoid nodes that make a choice in some shade of gray $[-1, 1]$ and the outputs move onto the next network layers. A purely feed-forward network, where there are no backward arcs, can be trained efficiently using back-propagation\cite{rumelhart2002learning} where classification errors are used to adjust the network weights backwards layer by layer.

In large NNs, such as those used in image processing, there can be a failure to generalize due to over fitting of the very large number of weights. One approach is to use a middle ``coding'' layer that is relatively small that forces the network to learn the key generalizations\cite{hinton2006reducing}. Recently, the so-called ``dropout'' algorithm has been developed that trains only subsets of the network on each example and this helps generalization too\cite{srivastava2014dropout}\bgref.

Closely related to NNs are logistic belief networks, where the nodes switch from 0 to 1 as a function of the probability of the weighted inputs. Hinton, Osindero, and Teh\cite{hinton2006fast} present a particular form of a multilayer belief network where the initial layers are feed-forward and the final two layers are interconnected in such a way to form an associative memory. An efficient training algorithm is developed that trains the individual layers using a greedy algorithm, and then refines the weights for the whole network. For a standard handwriting recognition benchmark (the MNIST database of handwritten digits) the error rate was 1.25\% which was better than that obtained by other standard machine learning techniques (SVM was second best at 1.4\%). However, if you train a standard NN using slight perturbations of the training data, i.e.~moving pixels around a bit, error rates as low as 0.4\% have been reported as of 2006. Table~1 of the reference shows some nice comparative data on methods and error rates.

Krizhevsky, Sutskever, and Hinton present their results from the ImageNet LSVRC-2010 and LSVRC-2012 contests\cite{krizhevsky2012imagenet}\footnote{The existence of standard data sets and contests has been very important in the development of machine learning algorithms.}. The goal was to classify images into categories, and the training data set has roughly 1,000 images in each of 1,000 categories for a total of about a million images. The authors trained a convolutional neural network having 60 million parameters using several optimizations to make the problem tractable (the input layers of CNNs are not fully connected, they ``focus'' on overlapping zones of the visual field much like biological systems). The resulting network, for LSVRC-2012, had an error rate of 15.3\% compared to the second-best entry's rate of 26.2\%.

There is substantial anecdotal evidence that NNs and SVMs are the most powerful classifiers if trained properly, and that is why their use is so widespread.  Hastie et al.\cite[Chapters 12 and 11]{hastie2009elements} contains chapters on SVMs and NNs. The classic text of Duda et al.~on pattern classification\cite{duda1999pattern} also covers NNs, genetic algorithms, and many other machine learning algorithms.

Finally, hidden Markov models\cite{rabiner1989tutorial} are transition networks where each transition is labeled with a probability of happening. They are common in natural language processing, but can also be applied to problems such as representing various biological (e.g.~regulatory) networks.

\subsection{General Tools and Applications}
\label{sec:tools}
The section describes some general tools and applications that appeared in the works due to their wide applicability. It opens with Google, which was somewhat surprising to the authors, but the company clearly has an impact on the thinking of data scientists. The section closes with several general tools, such as R and IPython.

\subsubsection{Google}

PageRank\cite{page1999pagerank} is an algorithm for ranking pages in web searches and was the first used by Google. It is an important example of applied computer science, where two good intuitions are combined in a mathematically rigorous way to produce an algorithm of high utility. The first intuition is that the importance of a page is proportional to the number of pages that link to it. Ultimately, the sum of the importances for all pages is one. The second, and more mathematically interesting is that there is a damping factor which is denoted $d.$ The idea is that a person will only wander so far (click) from a search result before getting bored and moving on to something else. In practice, $d \approx 0.85$\cite{brin1998anatomy}, and this $0 < d < 1$ helps to give rapid convergence.

Consider $N$ web pages where the PageRank of page $i$ is denoted $r_i$ and define $R^{T} = \{r_1, r_2, \ldots, r_N\}.$ Further define the matrix $M_{ij} = \delta_{ij} / L_j$, where $L_j$ is the number of outbound links from page $j$ and $\delta_{ij} = 1$ if pages $i,j$ are linked, otherwise it is zero. With the identity matrix $I, R$ is given in the steady state by
\begin{equation}
  R = (I - d M)^{-1} {1 - d \over N} I
\end{equation}
In practice, the solution is computed iteratively and converges quickly.

Conceptually, MapReduce\cite{dean2008mapreduce} transforms an input set $X$ of key:value pairs with keys in $K_1$ to an output set $Y$ of pairs with keys in $K_2$ using a three stage Map-Shuffle-Reduce process. The Map step applies a function to every element of $X$ producing an intermediate list $X’$ containing new pairs with keys in $K_2.$ This $X’$ is then Shuffled to group the values corresponding to a given key in $X’$ together so that they can then be Reduced using another function into the output $Y$. In the canonical example of counting the number of times a distinct word appears in a set of files, the elements of $K_1$ are filenames and $K_2$ contains words, the associated values are file contents and word counts.

If general, if $X$ and the post-shuffled $X’$ are distributed across many nodes, the map and reduce stages can be done in parallel on local data. Production implementations have many optimizations to deal with issues like load balancing, data positioning and replication, minimizing communications, and fault tolerance. PageRank can be formulated in a way that yields an efficient MapReduce implementation. In the context of data-intensive discovery, it is very common to combine MapReduce with machine learning and classification (Sec.~\ref{sec:ml}) to parallelize the processes.

The fact that a commercial enterprise is making such an impact on science is wonderful! However, we must add a note of caution: ``Big Data'' is not just the massive application of machine learning methods with large, blunderbuss, clusters; it is more subtle and widespread (Sec.~\ref{sec:scimeth}). Hadoop, an open implementation of MapReduce, was also cited by some. It should be noted that Google has largely moved onto systems such as BigTable\cite{chang2008bigtable}\bgref and Cloud Dataflow for storing and processing data (\verb|https://cloud.google.com/dataflow/|).

\subsubsection{General Tools}

A strong cluster of references emerged around tools, programming languages and methods for understanding data. These works represent a cross section of non-domain specific methods that researchers from a variety of disciplines are utilizing to process data to information to understanding.

Numerical Recipes\cite{press2007numerical} is the most widely used reference for numerical algorithms and it covers a broad range of topics from linear algebra to optimization. There have been multiple editions since 1986, and the most recent edition (2007) has been expanded to cover topics such as classification and inference. The series’ web site, {\tt www.nr.com}, considers itself one of the oldest pages on the Web, and provides paid access to all algorithms in various programming languages.

The R language\cite{team2012r} is one of the leading statistical programming languages, and was referenced a significant number of times in the {dataset}.  R was created as a free and open source implementation of the S statistical programming language with influences from Scheme.  R focuses on ease of use, tight integration with publication quality graphics and charts, data processing, and modular extensions to go beyond the core functionality.  It has its own mathematical formula expression language, like \LaTeX, and provides users convenient tools converting formulas into executable code. 

The IPython Notebook project\cite{perez2007ipython} (now Jupyter at {\tt www.jupyter.org}) is noteworthy as one of a few open source software toolkits for both programming and data analysis that is not a database, algorithm or programming language. Jupyter is an ``architecture for interactive computing and computational narratives in any programming language.'' It provides both a programming and documentation environment which ultimately allows for sharing of so-called narratives in an executable notebook, all available via the web. It is language agnostic; processing R, Python, Julia and provides basic workflow/reproducibility and collaboration capabilities. It is being used in a wide variety of scientific applications.

The Visual Display of Quantitative Information\cite{tufte1983visual} by Tufte is a seminal work on data visualization, with a focus that uses very powerful human perceptive systems that are not likely to be automated soon. The famous chart, of course, is Napoleon's foolish march and then retreat from Russia. The authors feel that, perhaps, all talks should be speeches and perhaps simply summarized in a few charts. Tukey in a 1977 work\cite{tukey1977exploratory} also emphasizes the use of graphs and tables to explore data.

Finally, Codd\cite{codd1970relational} introduced relational databases in a brilliant Tour de Force of computer science, coupling theory with practice. No longer were databases to be ad-hoc, they have a theory that could be used to make them better. This is at the core of all relational database systems, and Codd won the A.M.~Turing award in 1981 for his work.

An observation is the power of open source software. R, IPython, and Apache Hadoop which contains an implementation of MapReduce, are all available under various open source licenses. This allows the free use, inspection, and extension of the codes and greatly lowers barriers to entry, particularly for academic research purposes.

\subsection{Centrality of the Scientific Method}
\label{sec:scimeth}

One of the most cited influential works was {\it The Fourth Paradigm}\cite{hey2009fourth}, a collection of papers on data intensive scientific discovery produced by Microsoft in honor of Jim Gray, one of the first modern data scientists. The collection has had a catalytic effect based on the number of references, from researchers in a wide variety of fields. Another influential work is on the unreasonable effectiveness of data\cite{halevy2009unreasonable}, which is a nice play on the unreasonable effectiveness of mathematics. We must distinguish between {\it tools,} or instruments, and the scientific method. In the Fourth the argument is made that science has progressed from the 1.~empirical stage (observation-only), to the 2.~theory stage, and on to 3.~simulation based science, and finally 4.~big data science. It was at stage 2 that the scientific method became fully formed, and Newton deserves a lot of credit although Maxwell showed the raw power of theory to explain phenomena beyond human senses. The tools that Newton used were the calculus, which he had to invent, inclined planes, and dropping fruit. Now we use computers in stages 3 (theory) and 4 (observation). The scientific method stays the same, technology just allows better tools which begets deeper science and then new technology and tools.

There have also been claims that ``Big Data'' will eliminate science, we just need to use powerful methods to classify the data and from that we will know everything. The trouble is confusing classification, like botany, with science: predictive theories with bounded errors. Let us consider training a classifier to near Bayesian optimal. It could be a NN or a SVM, but the advantage with an SVM is that we can extract out the key support vectors, the prime $x_j$, and examine them. {\it Does this tell us anything?} The trouble is that if the experiment is changed, the support vectors will likely change too so where is the insight? Another take on this is by Breiman, in ``Statistical modeling: The two cultures''\cite{breiman2001statistical}, where he contrasts what is called classical statistical methods in this paper with algorithmic models. The comments associated with the paper are enlightening.

As a concrete example, it may be within current computing and algorithmic technology to infer the Maxwell Equations directly from data given knowledge of vector calculus. This would be a formidable achievement. Indeed, the kinematic laws of the double pendulum problem can be inferred using symbolic regression from observations\cite{schmidt2009distilling}. Latent in the Equations, however, is special relativity but it requires a mental shift to tease this out: specifically, Einstein's axiom that the speed of light is constant in all inertial reference frames. Making this brilliant leap seems hard to do by computing at this time. Perhaps we need a new Turing test, one not susceptible to linguistic parlor tricks: {\it given just the data and some fundamental theorems from analysis, discover special relativity and general relativity.}

A recent paper (2013) by V.~Dhar, ``Data science and prediction''\cite{dhar2013data}, defines {\it data science} as
\begin{quotation}
\ldots the study of generalizable extraction of knowledge from data.

A common epistemic requirement in assessing whether new knowledge is actionable for decision making is its predictive power, not just its ability to explain the past.
\end{quotation}
This view is entirely consistent with the scientific method, however it does not mean that the {\it way} scientists do science is fixed. Indeed, in the delightful book {\it Reinventing discovery}\cite{nielsen2012reinventing} M.~Nielsen argues that network effects in scientific communications and access to data will dramatically accelerate scientific discovery. This prediction is almost certainly true.

Finally, there were a few general references such as Han and Kamber\cite{han2011data} and the National Academy of Sciences report on the {\it Frontiers of Massive Data Analysis} \cite{national2013Frontiers}.

\section{Concluding Remarks}

\paragraph{Limitations} It must be noted that the competition was for efforts in the natural sciences and methodologies, and therefor references important to social sciences are underrepresented in this sample. Indeed, the social sciences are potentially one of the most impactful areas for big data and we encourage funders in these fields to run an investigator competition in this broad area.

As mentioned in the introduction, we asked applicants to tag works as papers, books, or resources. The matching algorithm works very well for papers and books, but not so well for data resources and software tools. The fundamental problem is that there is no commonly accepted way of citing resources unless there is an associated paper (e.g. IPython) or the authors are very specific about how to cite the tool (e.g. R). We are sure that if we went through the nearly 5,000 citations by hand, we would find more resources but we decided to stay with our deterministic, repeatable, methodology. Efforts to attach Digital Object Identifiers (DOIs) to resources are underway; however we believe one reason that articles and books are easier to reference is that they also have a standard, {\it human understandable}, way to identify themselves and not just some cryptic number.

\paragraph{Next steps} The concepts behind ``Big Data'' are not new, and go back to at least 1609 with Kepler's {\it Astronomia Nova} \cite{kepler1992astronomia}\bgref. The great, early, data scientist reduced Tycho Brahe's voluminous observational data into just three laws, the most famous probably being that bodies move in ellipses about a mass center\footnote{It may also be interesting to note that the publication of {\it Nova} was delayed by about 4 years, from 1605 to 1609, due to an intellectual property argument surrounding Mr. Brahe's data.}.

Our longer term plan is to perform further study of the influential works and to develop this preliminary paper into a review suitable for journal publication. At that time we will release the \textsc{Bib}\TeX\ file under a suitable Creative Commons license. The authors hope that a primary value of this work is in education.

\section*{Acknowledgments}
The authors thank the 1,095 applicants to the DDD Investigator Competition, the advisory panel, commentators on the initial arXiv paper (v1), and the members of the Moore Foundation Science Program and its (former) Chief Program Officer, Vicki Chandler. We would also like to thank Joshua Greenberg of the Alfred P.~Sloan Foundation for useful discussions and initial motivation for this paper. M.~Stalzer thanks the Aspen Center for Physics and the NSF Grant \#1066293 for hospitality during the editing of this paper.

\appendix

A total of 1,095 applications were received in late February 2014, containing 4,790 references. The author, title, etc.~of each reference was broken into a bag of words and these bags were assigned to {\it buckets} based on reference similarity using weighted word frequency by a sorting process. Specifically, the weight of a word $i$ that occurred $N_i$ times is
\begin{equation}
\ln N_w/N_i
\label{eqn:freq}
\end{equation}
where $N_w$ is the total number of unique words. In other words, words of lesser frequency carry somewhat more weight, leading to higher matching value. An obvious example is ``paradigm''. Words must be of length four or greater and appear twice or more; this eliminates stop words, e.g. ``and, the'', in English and words with no matching value, although it does throw out a bit of information.

References were sorted into the buckets based on the bucket's {\it signature.} A signature keeps the top eight words in a bucket by Equation~\ref{eqn:freq}, although when buckets are merged in the sorting process (see below) all words in both buckets are used to recompute the new merged signature so that signatures are refined over time.

The sorting algorithm is straightforward. Begin by assigning each reference to its own bucket and compute its signature. Take the first bucket and find a bucket whose signature matches to within a threshold; if there is a match, merge the two buckets. Repeat with the second bucket, and so on. The threshold is manually adjusted to produce strong groupings, with few extraneous references in each bucket. If the threshold is too high, nothing groups, and if it is too low, everything groups into one bucket. Some manual edits were done to clean up the buckets. Papers, books, and resources were treated separately (this is done by giving the type tags high weights).

Four or five words of each signature were then submitted to Google Scholar to get \textsc{Bib}\TeX\ entries. Google Scholar almost always listed the right work first, although the quality of the \textsc{Bib}\TeX\ entries is highly variable and often needed to be fixed. 

\newpage
\section*{A Note on the References}

Each reference contains a note on the number of times it was cited (``n 63''), and the number of applicants that self-identified in a field, such as computer science, that cited the reference (``CS 41''). Table~\ref{tab:fields} is the key to fields.

\begin{table}[h]
\centering
\begin{tabular}{l l}
Tag & Field \\
\hline
ACM & Applied and computational mathematics \\
AG & Agriculture \\
APHYS & Applied physics \\
ASPC & Aerospace \\
ASTRO & Astronomy and astrophysics \\
ASTROB & Astrobiology \\
ATMOS & Atmospheric science \\
BCS & Brain and cognitive science \\
BIO & Biology \\
BIOE & Bioengineering \\
BIOI & Bioinformatics \\
CBIO & Computational biology \\
CE & Computer engineering \\
CHEM & Chemistry \\
CHEME & Chemical engineering \\
CIVE & Civil engineering \\
CLI & Climate science \\
CS & Computer science \\
CSS & Computational social science \\
CSYS & Complex systems \\
DM & Data mining \\
EBIO & Evolutionary biology \\
ECO & Ecology \\
EE & Electrical engineering \\
ENGR & Engineering (general) \\
EPS & Earth and planetary science \\
ESE & Environmental science and engineering \\
EST & Energy science and technology \\
GENE & Genetics \\
GENOM & Genomics \\
GEOP & Geophysics \\
MATH & Mathematics \\
MATS & Materials science \\
MBIO & Biochemistry and molecular biophysics \\
ME & Mechanical engineering and solid mechanics \\
MED & Medicine \\
MMO & Marine microbiology and oceanography \\
NEURO & Neuroscience \\
OPSR & Operations research \\
PHYS & Physics \\
REMS & Remote sensing \\
SBIO & Systems biology \\
SML & Statistics and machine learning
\end{tabular}
\smallskip
\caption{Key to reference tags and fields.}
\label{tab:fields}
\end{table}

\bibliographystyle{IEEEtran}

\bibliography{ddd_iw_r5}

\end{document}